\documentclass[twocolumn,nofootinbib,preprintnumbers]{revtex4-1}
\usepackage{graphicx}
\usepackage{amsmath, amsthm, amssymb}
\usepackage{slashed}
\usepackage{url}
\usepackage[pdftex]{hyperref}
\usepackage{color}
\usepackage{bm}
\usepackage{mathtools}

\newcommand{\bea}{\begin{eqnarray}}
\newcommand{\eea}{\end{eqnarray}}

\begin{document}

\preprint{UCI-HEP-TR-2021-25}

\title{Annihilogenesis}
\author{Jason Arakawa}
\author{Arvind Rajaraman}
\author{Tim M.P. Tait}
\affiliation{Department of Physics and Astronomy, University of
California, Irvine, CA 92697-4575 USA}
\date{\today}
\begin{abstract}
We investigate a novel interplay between the decay and annihilation of a particle whose mass undergoes a large shift during a first order phase transition, leading to the particles becoming trapped in the false vacuum and enhancing their annihilation rates as the bubbles of true vacuum expand. This opens up a large region of the parameter space where annihilations can be important. We apply this scenario to baryogenesis, where we 
find that annihilations can be enhanced enough to generate the requires baryon asymmetry even for relatively tiny annihilation cross sections with modest CP asymmetries. 
\end{abstract}

\maketitle

\section{Introduction}
\label{sec:intro}

Cosmological phase transitions represent a dramatic change in the properties of the Universe over a relatively short period of its history, and
may play an important role in our understanding of conditions today.  
Already in the Standard Model (SM), the QCD phase transition at which \(SU(3)\) color confines \cite{OLIVE1981483}
is thought to separate a phase dominated by free quarks and gluons from one where the relevant degrees of freedom are baryons,
and the electroweak (EW) phase transition demarcates a period where the electroweak gauge symmetry is exact from one in which
the weak bosons and SM fermions have non-zero masses.
In the context of physics beyond the Standard Model (BSM), first order phase transitions (FOPTs) are frequently invoked to
catalyze interesting dynamics.
For instance, the interactions between the thermal bath and the expanding bubbles of true vacuum typically present in a FOPT play a central role in mechanisms such as 
electroweak baryogenesis \cite{Kuzmin:1985mm, Carena:1996wj, Cohen:1993nk, Croon_2018}, where the FOPT realizes a departure from thermal equilibrium -- 
one of the three necessary conditions required for baryogenesis \cite{Sakharov:1967dj}.

In this article, we investigate a FOPT producing a large shift in the mass of a BSM particle \(\chi\), and explore how this leads to an interesting interplay between
the role of \(\chi\) decay and \(\chi \chi\) annihilation into SM particles during the FOPT itself.  After bubbles of the true vacuum nucleate, the $\chi$ mass can be
radically different inside and outside.  As the bubbles expand and collide
(using the terminology of Ref.~\cite{Asadi:2021pwo}), segmented ``pockets" of unbroken phase remain, and
experience contraction as the bubbles grow to fill the entire Universe. 
While this happens, $\chi$ particles in the pockets reflect off the bubble walls due to the 
large \(M^{\rm{in}}_{\chi}/T\) in the broken phase and as a result are trapped in the pockets,``squeezing" them together. 

We focus on the interplay between decay and annihilation processes during the pocket collapse, and analyze under which situations one or the other
can become the dominant mechanism depleting the particles. 
Generically, one would expect that decays, if allowed, would dominate over annihilation processes such that the depletion is governed by the decays alone.
However, as they are squeezed inside a contracting pocket, the particle densities may grow large enough to provide enough enhancement of the annihilation
rate that a significant number of \(\chi\) annihilate rather than decay, even for large decay widths. 
We find that depending on the parameters of the theory, the decay and annihilation can compete or be relevant at different times during the phase transition. 
This mechanism thus provides a novel relationship between the depletion processes, and can open up large regions of the parameter space in which annihilation can become important
or even dominate over decay.

As a specific application, we apply this scenario to baryogenesis. Interference between tree-level and loop-level diagrams can lead to a CP asymmetry in both decay and annihilation,
and even if the decay and annihilation processes are  governed by the same couplings (which they need not be), there are additional contributions 
to a CP asymmetry from the annihilation processes, and therefore the asymmetries generated by decay and annihilation are not constrained to be the same. 
We work in a generic framework, in which a FOPT traps the particles to decay or annihilate in the pockets of unbroken phase.
Previous related work \cite{Baldes:2021vyz} has examined baryogenesis in a similar context with 
relativistic bubble walls, but under the assumption that the effect of reflection off of the bubble walls is negligible.  

Recent studies have investigated similar ideas in the context of dark matter (DM), 
and how the DM relic abundance may be set by interactions with non-relativistic bubble walls via a ``filtering'' effect \cite{Baker:2019ndr,Chway:2019kft}, 
leading to an exponentially suppressed abundance of DM inside the bubbles. 
Other work has focused on the fate of the DM particles that reflect off the bubble wall and are trapped in the unbroken phase.  The particles trapped in the pockets are 
eventually ``squeezed" together, leading to a number of possible outcomes, depending on the specifics of their interactions. 
The squeezing could enhance their annihilation rate, 
which may determine the DM relic density \cite{Asadi:2021pwo}, or increase the density sufficiently enough to create compact objects such as 
primordial black holes or Fermi-balls \cite{Baker:2021nyl,Marfatia:2021twj}, which may themselves play the role of dark matter in the Universe today.

Our paper is organized as follows. Section~\ref{sec:mech} introduces the general framework and outlines the relevant features of a first order phase transition. Section~\ref{sec:squeeze} examines the interplay between decay and squeezed annihilation via the Boltzmann equation, and determines whether decay or annihilation is the dominant depletion process. Section~\ref{sec:asymmetry} discusses the asymmetry that is generated for different amounts of decay and annihilation. Section~\ref{sec:GW} shows the gravitational wave spectrum that could be produced within this general framework. We reserve section~\ref{sec:conclusions} for our conclusions and outlook. 

\section{General Scenario}
\label{sec:mech}

\begin{figure*}
    \centering
    \includegraphics[width = .49\textwidth]{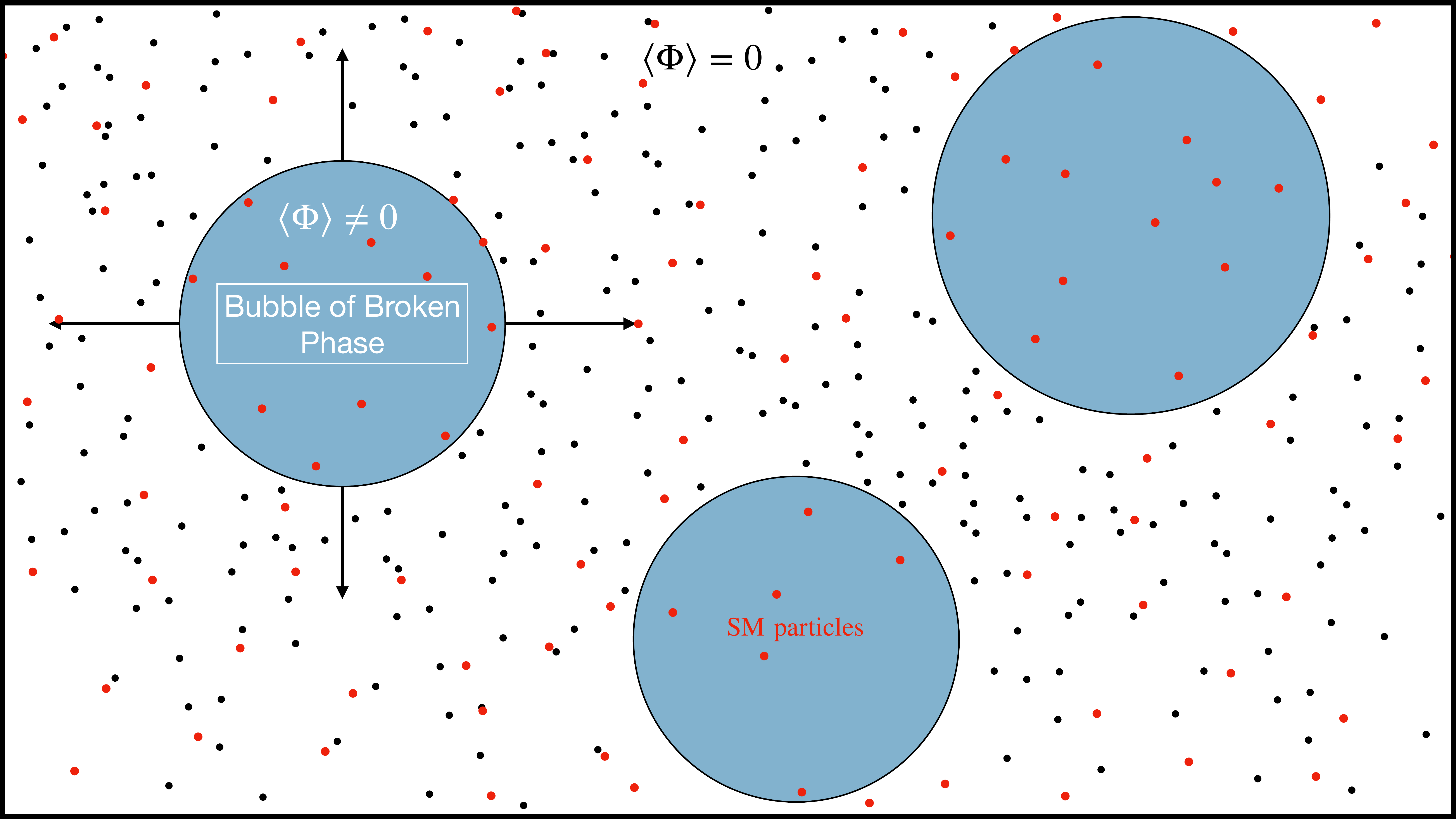}
    \includegraphics[width = .49\textwidth]{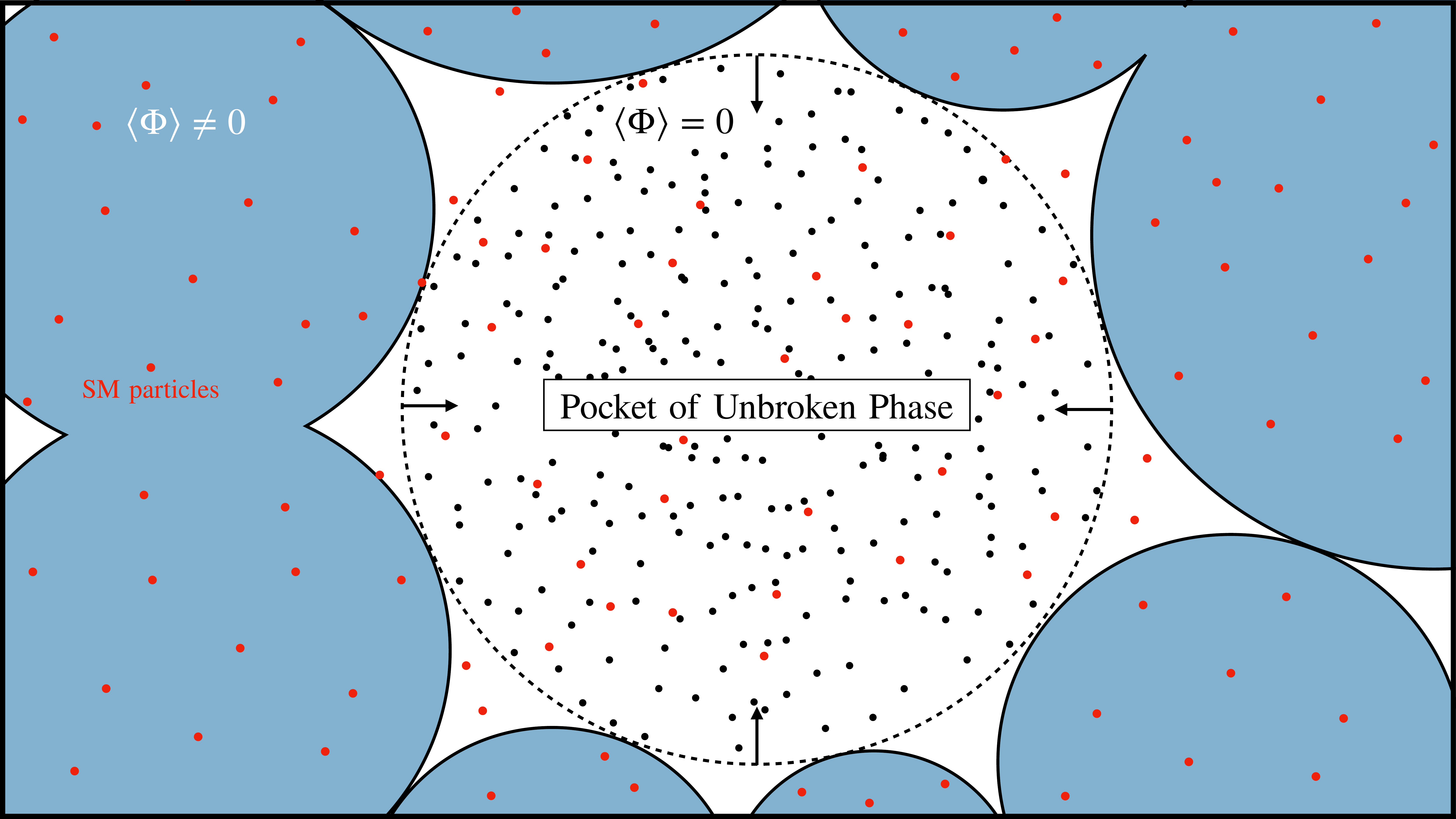}
    \caption{A cartoon depiction of the bubbles nucleating and expanding (left). As these bubbles collide, they create contracting pockets of unbroken phase, 
    which trap and squeeze \(\chi\) particles (black), enhancing their density (right), whereas SM particles (red) are able to traverse unimpeded. 
    We approximate the contracting pockets to be spherical.}
    \label{fig:SqueezeDiagram}
\end{figure*}

We consider a scenario where a fermion \(\chi\) is coupled to a complex scalar \(\Phi\) described by the Lagrangian,
\begin{align}
\label{eq:chiL}
    \mathcal{L} = \overline{\chi}(i \slashed{D}) \chi- y\Phi \overline{\chi}\chi+ \rm{h.c.} - V(\Phi)
\end{align}
where we assume for simplicity that both \(\chi\) and \(\Phi\) are SM singlets.
In order to consider a wide spectrum of scenarios,
we assume that there are couplings which mediate both the decay and annihilation of \(\chi\) into appropriate SM states, but do not specify their specific form. 

To successfully realize baryogenesis, there must be a source of CP violation in either the new sector itself or couplings between the new sector and the SM. 
If there are multiple flavors of  \(\chi_i\), this CP violation may come directly from the \(\Phi\) couplings,
\begin{align}
    \mathcal{L}_{CP} \supset y_{ij}\Phi \overline{\chi}_i\chi_j + y^*_{ij}\Phi^* \overline{\chi}_j\chi_i
\end{align}
which could generate CP violation via vertex corrections, self-energy corrections, and other loop level processes involving \(\Phi\). 
For now, we consider \(\chi\) to be the lightest species of the multiple generations, with any heavier states showing up only inside these loop-level processes. 
The Sakharov conditions additionally
require the presence of C and baryon number violation, which constrains the space of the generic couplings.

We assume that the thermal potential for \(\Phi\) is such that at some temperature in the early Universe it undergoes a first order phase transition, nucleating bubbles in the process.
The form of Eq.~(\ref{eq:chiL}) is such that
 at temperatures above the \(\Phi\) phase transition, the \(\chi\) have zero tree level mass. After the \(\Phi\) phase transition, the \(\chi\) are massive inside the bubbles of broken phase 
 (the phase where \(\Phi\) has a vev) and their mass is \(M^{\text{in}}_{\chi}= y \langle \Phi \rangle\). If the ratio \(M^{\rm{in}}_{\chi}/T \gg 1\), 
 then only the high momentum modes of \(\chi\) can penetrate the bubble wall, resulting in a large number of the \(\chi\) particles being trapped in the unbroken phase. 
Altogether this amounts to an out-of-equilibrium process with C, CP, and baryon number violation: all of the necessary ingredients to generate a baryon asymmetry. 

Throughout the remainder of the paper, we will use terminology introduced in Ref.~\cite{Asadi:2021pwo}. 
The regions we refer to as bubbles are the usual FOPT bubbles that nucleate and expand. 
As these bubbles collide, segmented regions of unbroken phase contract, which we refer to as ``pockets''. 

\begin{figure}[t]
    \centering
    \includegraphics[width = \linewidth]{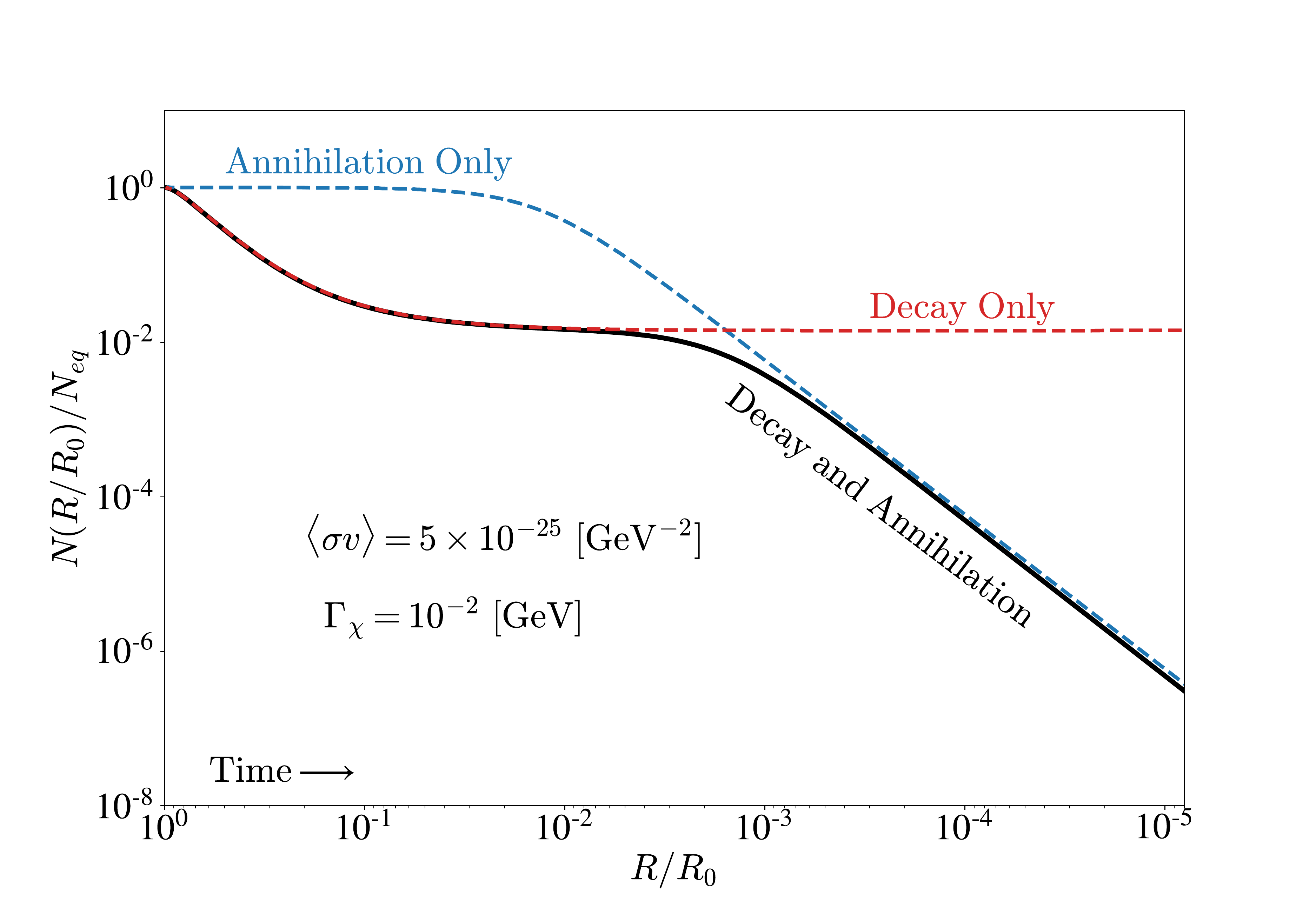}
    \caption{Example of \(\chi\) evolution during the pocket collapse, showing the fraction of the \(\chi\) that are present at different radii as the pocket collapses 
    in cases where only decay processes are present (red), only annihilation processes are present (blue), and both decay and annihilation processes are present (black).}
    \label{fig:GeneralBehavior}
\end{figure}

As the bubbles nucleate and expand, the particles with insufficient kinetic energy to enter the broken phase reflect off the bubble wall. 
The bubbles eventually collide, and isolated pockets of unbroken phase are left to contract (see Fig.\,\ref{fig:SqueezeDiagram}). 
During this pocket collapse, both decay and annihilation processes can both be important in the depletion of \(\chi\) as shown in Fig.\,\ref{fig:GeneralBehavior}
for a specific choice of parameters. 
Although the tree-level mass is zero in the unbroken phase, the thermal mass can allow the decays to become kinematically accessible. 
If the decay lifetime of \(\chi\) is shorter than the collapse time, then the \(\chi\) will start depleting via decays. 
Simultaneously, the pocket contracts, enhancing the annihilation processes as the pocket squeezing increases the density of the leftover \(\chi\). 
Whether the decay or annihilation processes dominate in depleting the \(\chi\) abundance depends on the relationship between the decay width, \(\Gamma_{\chi}\), 
annihilation cross section, \(\langle \sigma v \rangle\), and the pocket collapse rate.

\subsection{Phase Transition}

In general, the properties of the phase transition are largely governed by the potential \(V(\phi)\) (including thermal corrections).
Typically (even in the absence at tree-level), a cubic term arises from the high temperature expansion of the thermal loop corrections. 
This creates a barrier between the two minima, inducing a first order phase transition. 
One can generically write the finite temperature potential as \cite{Kehayias:2009tn}
\begin{align}
    V(\phi,T) = D (T^2 - T_0^2) \phi^2 - ET\phi^3 + \frac{\lambda(T)}{4} \phi^4
\end{align}
with \(D,~E,~\rm{and}~\lambda(T)\) determined by a combination of tree level potential parameters and both thermal and
zero-temperature loop corrections.
These parameters determine the critical temperature, \(T_c\), nucleation temperature, \(T_n\), and the strength of the phase transition,
with a strong first-order phase transition satisfying the condition \(\langle \phi\rangle/T_c \gg 1\).
In a specific theory, the coefficients $D$, $E$, and $\lambda$ can be computed, but rather than get distracted by these specific details, we treat them as
parameters that we can freely tune to realize a FOPT with various properties.
We also assume that the phase transition completes quickly enough that the temperature can be treated as a constant throughout its progress.

The typical initial size of the pockets will directly influence the dynamics and timescales that govern the \(\chi\) particles during the pocket collapse. 
The average number of bubbles that nucleate per Hubble volume scales as \(N_b \sim \beta_H^3\), where \(\beta_H\), 
is typically of order \(\mathcal{O}(10-10^4)\) for strongly FOPTs \cite{Baldes:2021vyz}, but could be as large as \(\mathcal{O}(10^{11})\)\cite{Marfatia:2020bcs}. 
This determines the initial size of the pockets by specifying the number density of bubbles that nucleate, \(n_{b} \sim \beta_H^3  H^3\), 
and the distance between bubble centers scales as  \(d_b \sim n_b^{-1/3} \sim R_H/\beta_H\) \cite{Megevand:2017vtb}. 
We consider both small and large initial pocket sizes by exploring two 
representative choices of the initial radii, \(R_0 = R_H\), and \(R_0 = 5\times 10^{-6}\, R_H\).

The bubble wall velocity \(v_w\) influences the rate at which the pockets contract. 
\(v_w\) can be estimated from thermodynamic arguments, 
but this neglects the pressure exerted by \(\chi\) particles reflecting off the wall, which could slow down the bubble 
expansion considerably~\cite{Asadi:2021pwo,Baker:2021nyl,Marfatia:2021twj}.  We consider both relativistic and non-relativistic wall velocities, where the larger the wall velocity, 
the larger the mass needs to be in the broken phase in order to trap \(\chi\) in the pockets. 
We choose  \(v_w = 0.9\), \(M^{\rm{in}}_{\chi}/T=10^2\) and \(v_w = 10^{-3}\), \(M^{\rm{in}}_{\chi}/T=10\) as two representative
examples, and assume that the wall velocity is approximately constant throughout the phase transition.

\section{Decays and Squeezed Annihilation}
\label{sec:squeeze}

Throughout the process of collapse, interactions with the thermal bath generate a thermal mass for \(\chi\)
of order \(\Pi_{\chi}^2 \sim g^2 T^2\) (where \(g\) represents a generic coupling to the thermal bath).
The particles that are trapped in these pockets are subsequently squeezed together and effectively obtain a Casimir mass, 
\(M_{\chi}^{\rm{cas}} \sim 1/R\), where \(R\) is the pocket radius.  This Casimir energy is an inherently quantum mechanical effect, due to the \(\chi\)
wave-functions' energies being bounded from below because of the size of the pocket they are confined within. 
For sufficiently confined \(\chi\), this mass can allow \(\chi\) to rapidly decay even when its tree level/thermal mass would otherwise forbid it from
doing so.  We denote the decay width of \(\chi\)  in the pocket as \(\Gamma_{\chi}\). 
We further assume that \(\chi \chi \) is also able to annihilate into SM final states with an annihilation cross section \(\langle \sigma v \rangle\).

As the pocket radius decreases to \(R \sim 1/M^{\rm{in}}_{\chi}\), the Casimir energy overcomes the potential energy barrier between the unbroken and broken phases, 
and the remaining abundance of \(\chi\) is forced into the bubbles where they eventually decay away. 
In Fig.\,\ref{fig:GeneralBehavior}, we show an example of the evolution of the number of \(\chi\) throughout pocket collapse.
Decays start immediately, governing the abundance early on in the phase transition. 
As the radius shrinks, there is less time left in the phase transition to allow for decays to occur, and the abundance due to decays flattens. 
However, at very small radii, the density of \(\chi\) increases enough to enhance the annihilation rate appreciably, allowing for a new depletion process
to become relevant.

\subsection{Boltzmann Equation}

\begin{figure*}
    \centering
    \includegraphics[width =\textwidth]{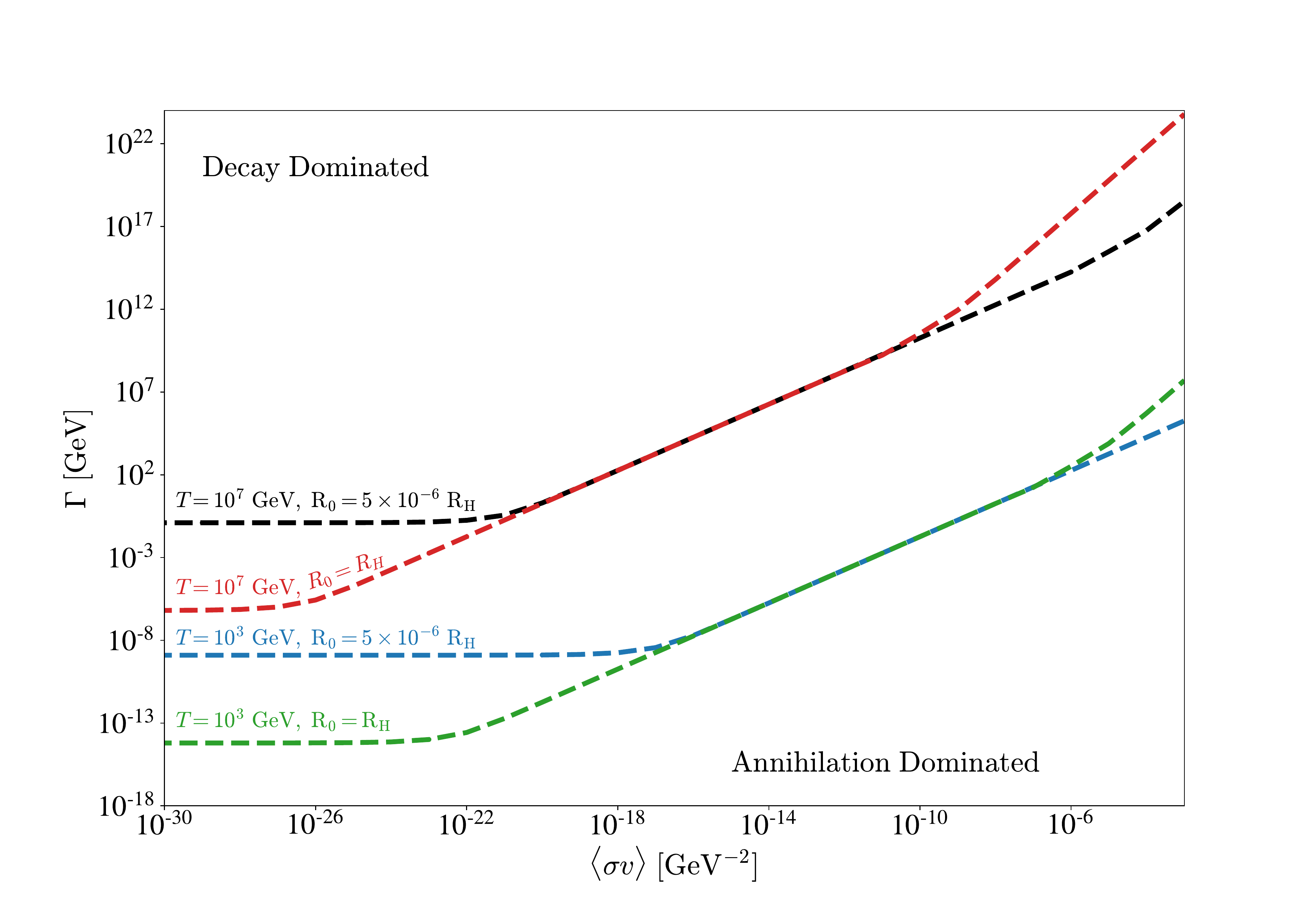}
    \caption{Contours in the \(\Gamma\)-\(\langle \sigma v \rangle\) plane corresponding to the points where \( \chi \)
    is depleted by an equal number of decays and annihilations during the pocket collapse.  The four curves correspond to combinations of
    temperature \(T = 10^7\) GeV and initial radius \(R_0 = 5\times 10^{-6} R_H\) (black); \(T = 10^7\) GeV and \(R_0 = R_H\) (red); 
    \(T = 10^3\) GeV and \(R_0 = 5\times 10^{-6} R_H\) (blue); and \(T = 10^3\) GeV and \(R_0 = R_H\) (green).}
    \label{fig:Ann_Dec_Half}
\end{figure*}

We track the abundance of \(\chi\) throughout the pocket collapse by solving a Boltzmann equation 
for the number density
of \(\chi\) confined inside the contracting pocket 
\begin{align}
    \frac{dn_{\chi}}{dt} + 3\frac{\dot{R}}{R}n_{\chi} = &-\langle\sigma v\rangle (n_{\chi}^2 - n_{\rm{eq}}^2)\nonumber\\
    &-\Gamma_{\chi} (n_{\chi} - n_{\rm{eq}}) .
\end{align}
We assume that \(\chi\) have sufficiently strong interactions with the SM plasma 
that they have their equilibrium abundance at the beginning of the phase transition,
and we approximate the pocket to be spherical with a constant wall velocity, \(v_w\). 
The Boltzmann equation can be recast into an equation differential in the radius of the pocket 
by making use of the relation \(\frac{dn_{\chi}}{dt}=\frac{dn_{\chi}}{dR}\frac{dR}{dt} = -v_w \frac{dn_{\chi}}{dR} \),
\begin{align}
    -v_w \frac{dn_{\chi}}{dR} - 3\frac{v_w}{R}n_{\chi} = &-\langle\sigma v\rangle (n_{\chi}^2 - n_{\rm{eq}}^2)\nonumber\\
    &-\Gamma_{\chi} (n_{\chi} - n_{\rm{eq}}) .
    \label{Boltz}
\end{align}
For a generic point in parameter space, both annihilation and decay may be significant. 

The total number of \(\chi\) inside the pocket is \(N_{\chi} = 4\pi R^3 n_{\chi}/3\), for which:
\begin{align}
    \frac{dN_{\chi}}{dR} &= -\frac{4 \pi R^3}{3 v_w} \bigg(\langle\sigma v\rangle (n_{\chi}^2 - n_{\rm{eq}}^2)
    +\Gamma_{\chi} (n_{\chi} - n_{\rm{eq}})\bigg) .
\end{align}
To determine the dominant process responsible for depleting the abundance inside the pocket, we compute 
the fraction of the depletion that was from annihilation, \(f_{A} = \Delta N_{\rm{annihilation}}/\Delta N_{\rm{total}}\), 
by comparing the integral of the corresponding terms in the Boltzmann equation,
\begin{align}
    f_{A} &= \frac{1}{\Delta N_{\rm{total}}}\int dN_{\rm{annihilation}}\\
    &= \frac{\displaystyle\int_{R_0}^{1/M} R^3 dR ~\langle\sigma v\rangle (n_{\chi}^2 - n_{\rm{eq}}^2)}{\displaystyle\int_{R_0}^{1/M} R^3 dR~ \bigg(\langle\sigma v\rangle (n_{\chi}^2 - n_{\rm{eq}}^2)+\Gamma_{\chi} (n_{\chi} - n_{\rm{eq}})\bigg)} . \nonumber
\end{align}

In Fig.~\ref{fig:Ann_Dec_Half} we show contours in the plane of \(\Gamma\)-\(\langle \sigma v \rangle\) corresponding to equal depletion by decay and annihilation
for \(v_w = 10^{-3}\), \(M^{\rm{in}}_{\chi}/T=10\) and
for four different combinations of the initial pocket size \( R_0 \) and the temperature $T$ at which the phase transition takes place.
Generally as expected, larger widths correspond to decay domination, and larger cross sections to annihilation cross section, with the boundary of \( f_A = 1/2 \)
determined by the temperature, which controls the initial density of \( \chi \) and thus the rate of annihlation.
However, there is a flattening at low \(\langle \sigma v \rangle\) which occurs when the decay and annihilation processes are operating during different times. 
In this case, the decays start immediately and the contour of \(f_A = 1/2\) corresponds to the point where half of the initial abundance inside the pocket decays before 
the time where squeezing becomes sufficient that annihilations turn on and deplete the rest of the abundance. 

For a phase transition with different \( v_w \), the dominant difference is through the explicit dependence in equation~(\ref{Boltz}), which
can be rescaled such that the quantities driving the evolution of \( n_\chi\) are \( \Gamma_{\chi}  / v_w \) and \(\langle \sigma v \rangle / v_w \).
For larger \(v_w\), in order to keep the \(\chi\) confined to the pockets, the phase transition must also have a larger value of \(M^{\rm{in}}_{\chi}/T\) which
further implies that the \(\chi\) reach sufficient Casimir energy to escape the pockets at a smaller pocket radius, and thus there is a slightly longer
period for decay and annihilation to operate.   For the relativistic wall velocity case we consider with \(v_w = 0.9\) and \(M^{\rm{in}}_{\chi}/T=10^2\),
this second effect is numerically unimportant, and the contours of fixed $f_A$ are very close to 
those shown in Figure~\ref{fig:Ann_Dec_Half} with appropriate rescaling by \( v_w \).

\begin{figure*}
    \centering
    \includegraphics[width = .49\textwidth]{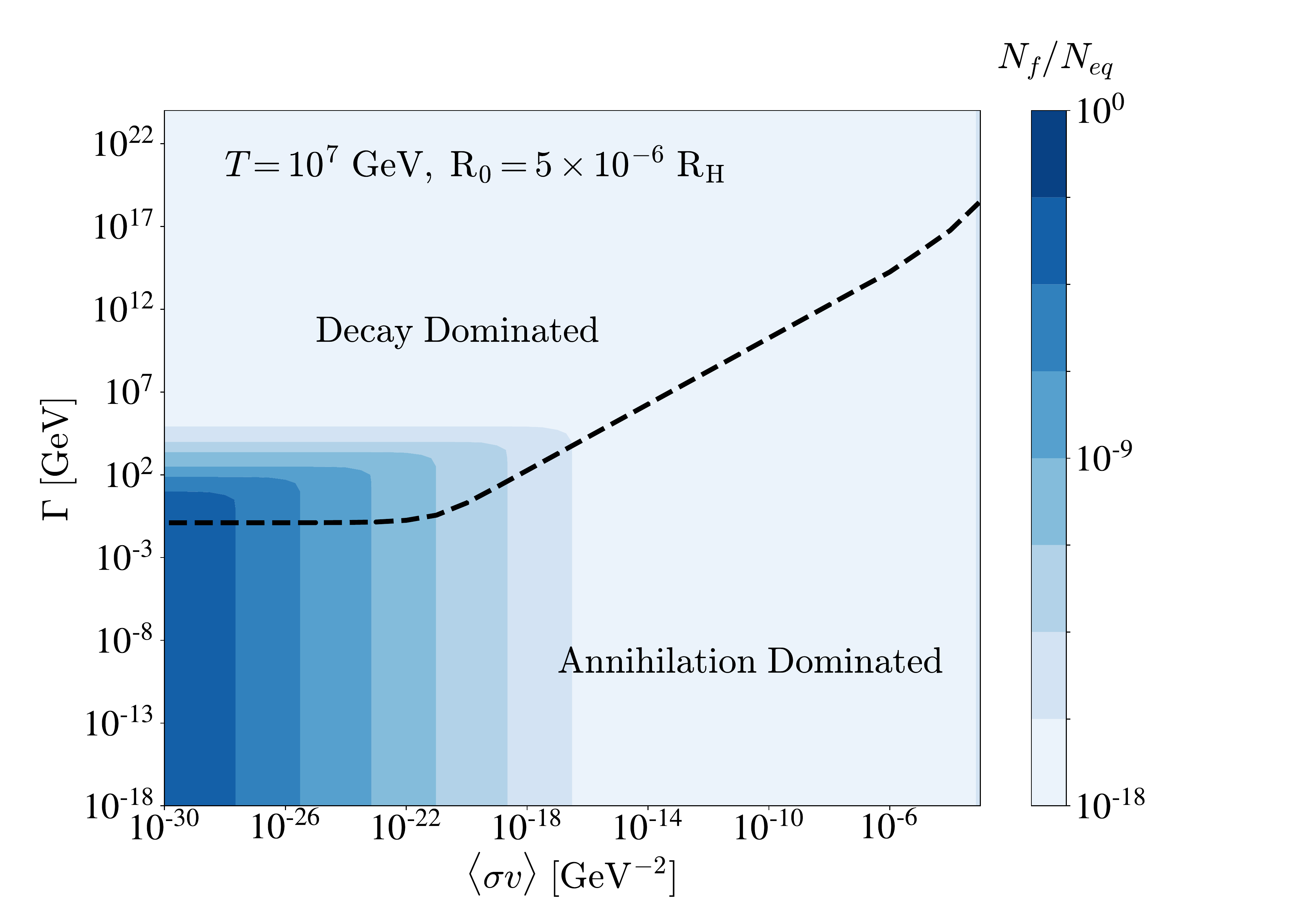}
    \includegraphics[width =
    .49\textwidth]{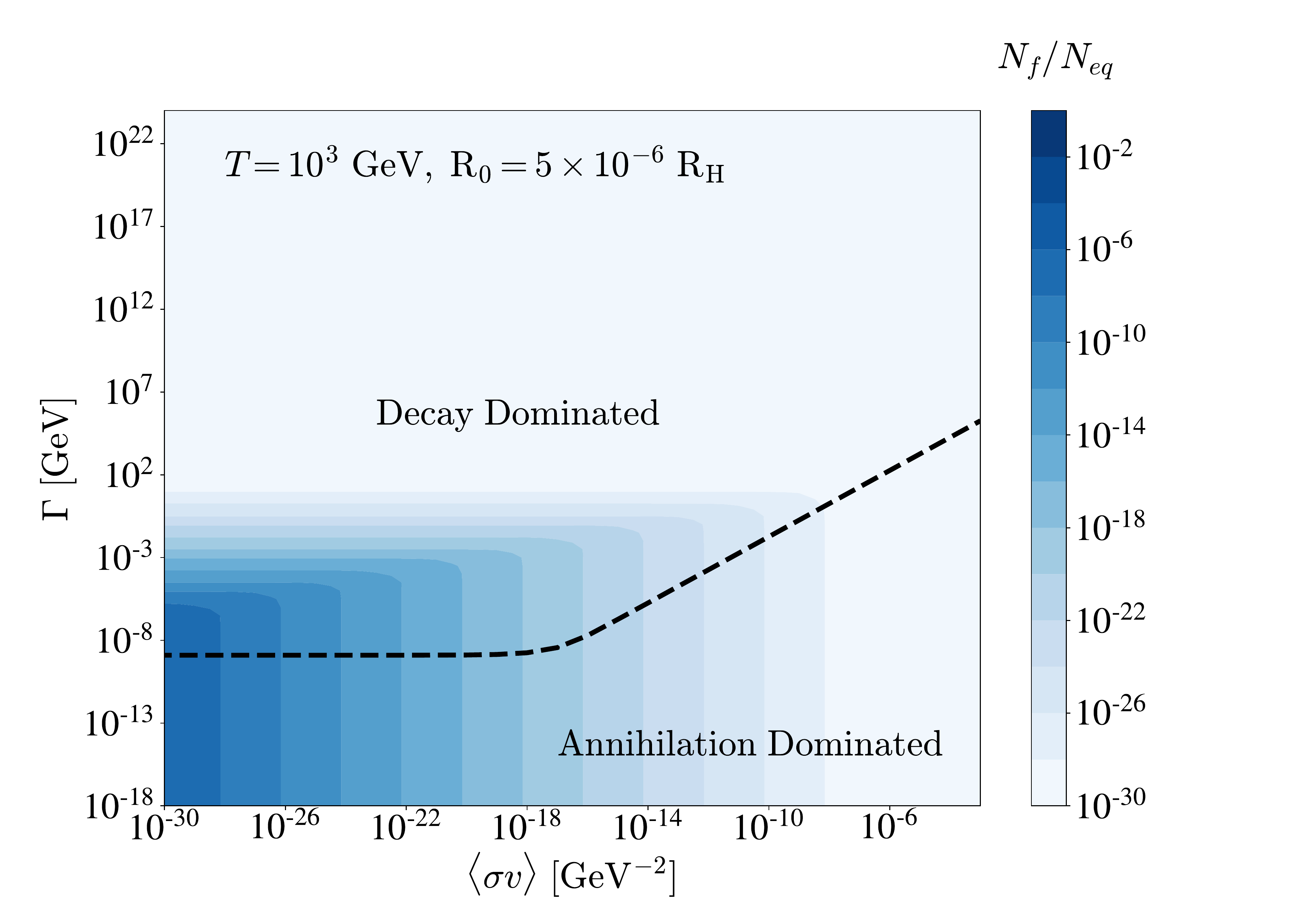}
    \includegraphics[width = .49\textwidth]{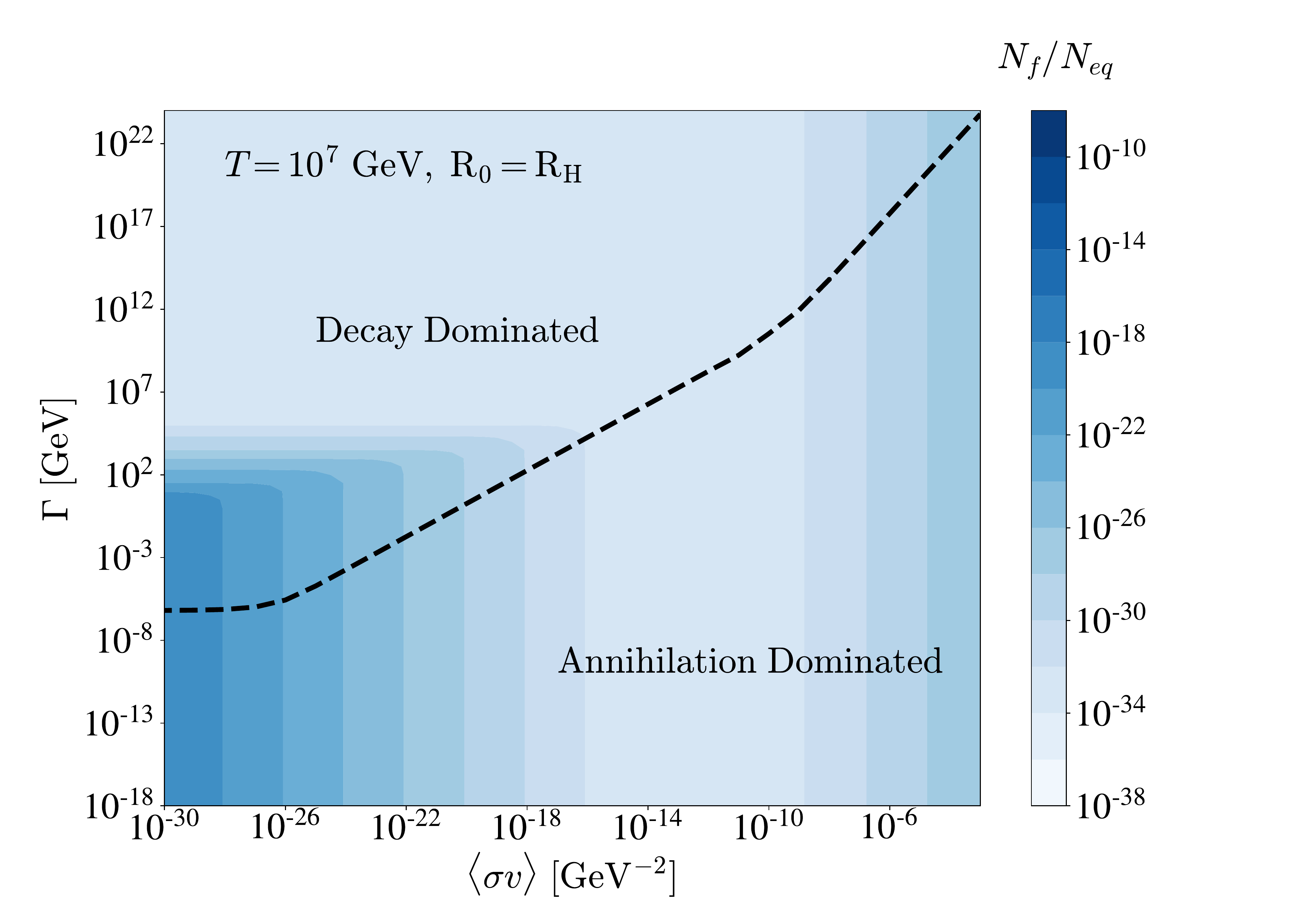}
    \includegraphics[width = .49\textwidth]{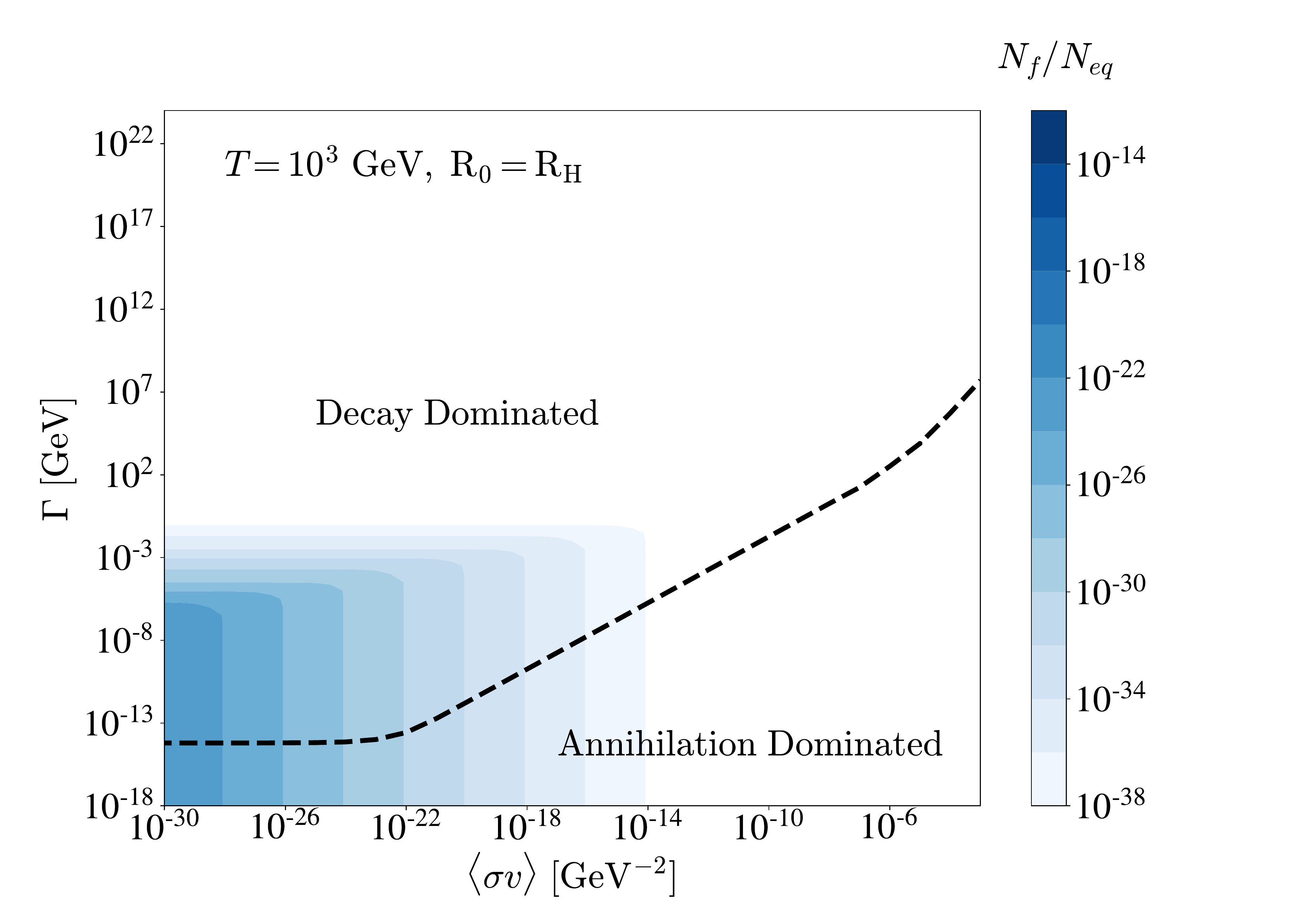}
    \caption{Contours in the \(\Gamma\)-\(\langle \sigma v \rangle\) plane indicating the fraction of the initial abundance that remains in the unbroken phase 
    when the pocket radius reaches \(R = 1/M^{\rm{in}}_{\chi}\), for the same four combinations of \( T \) and \( R_0 \) as in Figure~\ref{fig:Ann_Dec_Half}. 
    The black contours correspond to points shown in Figure~\ref{fig:Ann_Dec_Half} where \( f_A = 1/2 \).}
    \label{fig:Ann_Dec_Frac}
\end{figure*}

Fig.\,\ref{fig:Ann_Dec_Frac} displays for each of the parameter sets shown in Fig.\,\ref{fig:Ann_Dec_Half} the total depletion in the plane of \(\Gamma\)-\(\langle \sigma v \rangle\).
Depletion in very efficient in most of the plane, but in regions with both very small decay widths and annihilation cross sections, there could be a population
of \( \chi \) that survive the pocket collapse.

\section{Application to Baryogenesis}
\label{sec:asymmetry}

We consider an application of these results to baryogenesis, continuing to work in a generic framework in which \( \chi \) particles can both decay and annihilate into
SM states, and including the possibility of CP violation (as well as C and baryon-number violation)
being present in both processes.  
As noted above, the specific interactions mediating \(\chi\) decay or annihilation
may be different (and thus have intrinsically different CP violation).  Even if the underlying source of CP violation is the same for both processes, they will still
generically manifest themselves differently, because of different topologies of loop diagrams that contribute.

We parameterize the asymmetries present in the decay and annihilation processes as
\(\epsilon_{D}\) and \(\epsilon_A\), respectively:
\begin{align}
\epsilon_{D} &\equiv \frac{\sum_{\alpha}\left[\Gamma\left(\chi \rightarrow \rm{SM} \right)-\overline{\Gamma}\left(\chi \rightarrow \rm{SM}\right)\right]}
{\sum_{\alpha}\left[\Gamma\left(\chi \rightarrow \rm{SM}\right)+\overline{\Gamma}\left(\chi \rightarrow \rm{SM}\right)\right]} \\
\epsilon_{A} &\equiv \frac{\sum_{\alpha}\left[\sigma\left(\chi \chi \rightarrow \rm{SM\, SM}\right)-\overline{\sigma}\left(\chi \chi \rightarrow \rm{SM\, SM}\right)\right]}
{\sum_{\alpha}\left[\sigma\left(\chi \chi \rightarrow \rm{SM\, SM}\right)+\overline{\sigma}\left(\chi \chi \rightarrow \rm{SM\, SM}\right)\right]} .
\end{align}
We assume, as is typically the case, that \( \epsilon_{D}, \epsilon_{A} \ll 1 \).

The asymmetry generated by the combined decay and annihilation processes, \(\epsilon_{\rm{total}}\), 
is obtained by integrating the Boltzmann equation, keeping track of the fraction of \(\chi\) that annihilate versus decay:
\begin{align}
    \epsilon_{\rm{total}} =& ~\frac{1}{\Delta N_{\chi,\rm{total}}}\int \bigg(\epsilon_{A} \frac{dN_{\rm{ann}}}{dR}+ \epsilon_{D} \frac{dN_{\rm{decay}}}{dR}\bigg)dR \nonumber\\
    =& ~ \epsilon_{A} f_{A} + \epsilon_{D} (1-f_{A}) .
\end{align}
The resulting baryon asymmetry can be parameterized as:
\begin{align}
    Y_{\Delta B} = \Delta Y_{\chi}\epsilon_{\rm{total}}C = \frac{\Delta n_{\chi}(T)}{s(T)}\epsilon_{\rm{total}}C
\end{align}
where \(Y_i = n_i/s\), \(\Delta n_{\chi}\) is the total change in the number density of \(\chi\) during the pocket collapse, \(s(T)\) is the entropy density, and
\(C\) translates from the CP asymmetry present in the \(\chi\) depletion processes to the final asymmetry in baryons.
For example, if \(\chi\) depletion produces an asymmetry in lepton number that is subsequently 
converted into a baryon asymmetry via electroweak sphalerons (``leptogenesis"), \(C \simeq 12/37\) \cite{Davidson:2008bu}. 

\begin{figure*}
    \centering
    \includegraphics[width = .7\linewidth]{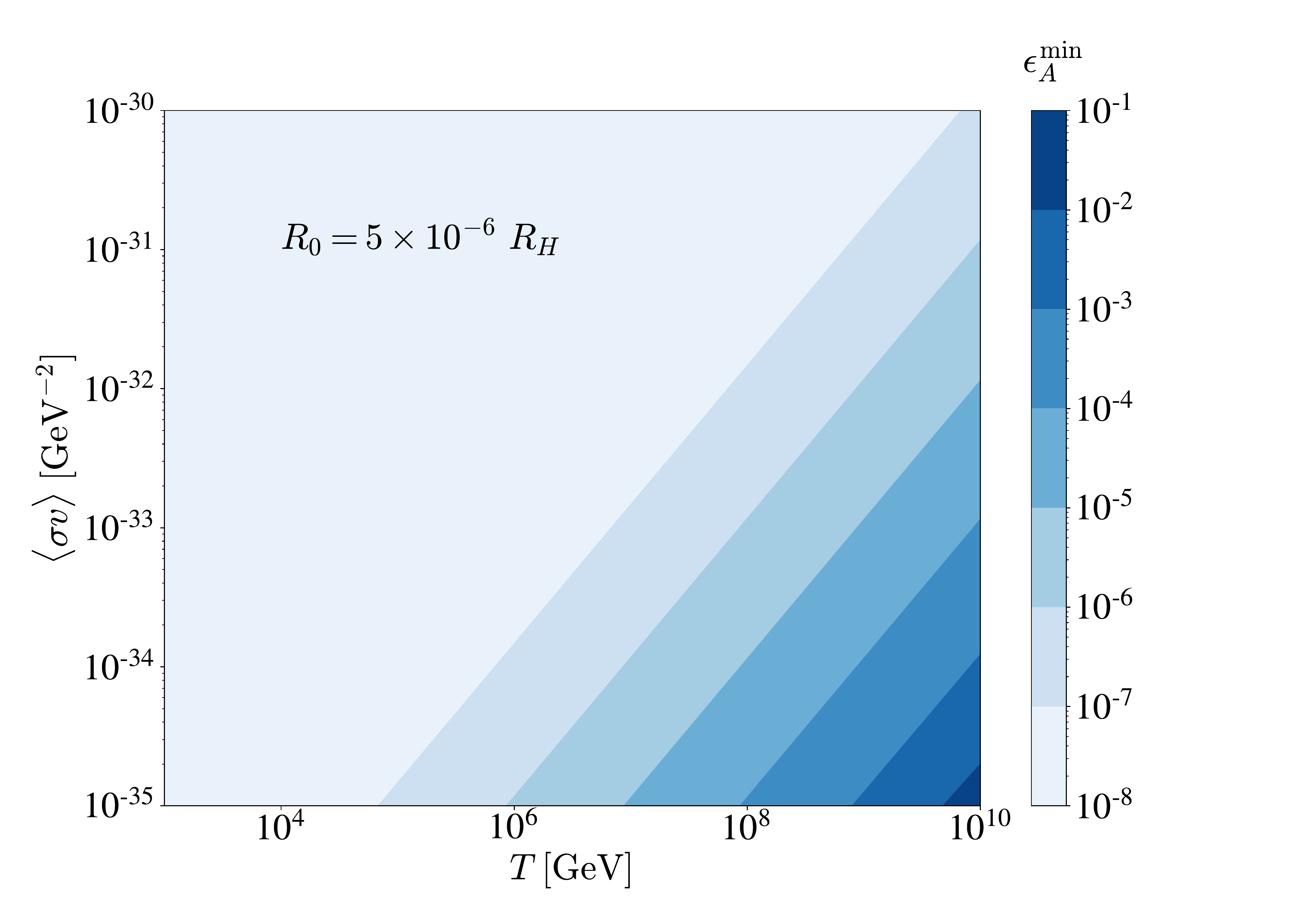}
    \caption{The minimum asymmetry needed to be generated by the annihilation processes in the \(\langle \sigma v \rangle - T \) plane, in the scenario 
    with  \(R_0 = 5 \times 10^{-6} R_H\), where annihilation dominates.}
    \label{fig:MinAsymmetry}
\end{figure*}

In Fig.~\ref{fig:MinAsymmetry}, we display the minimum \(\epsilon_A^{\rm{min}}\) in annihilations that is 
necessary to produced the observed baryon asymmetry, in the \(\langle \sigma v \rangle - T\) plane, for the annihilation-dominated phase transition with
 \(R_0 = 5 \times 10^{-6} R_H\) and \(C = 1\). Even for the tiny \(\langle \sigma v \rangle\) considered, there is sufficient enhancement
 provided by the squeezed annihilation rate in much of the parameter space to generate a sufficient baryon asymmetry
 such that only a rather modest \(\epsilon \sim \mathcal{O}(10^{-8})\) is needed.

\section{Gravitational Waves}
\label{sec:GW}

\begin{figure*}[t]
    \includegraphics[width = .49\linewidth]{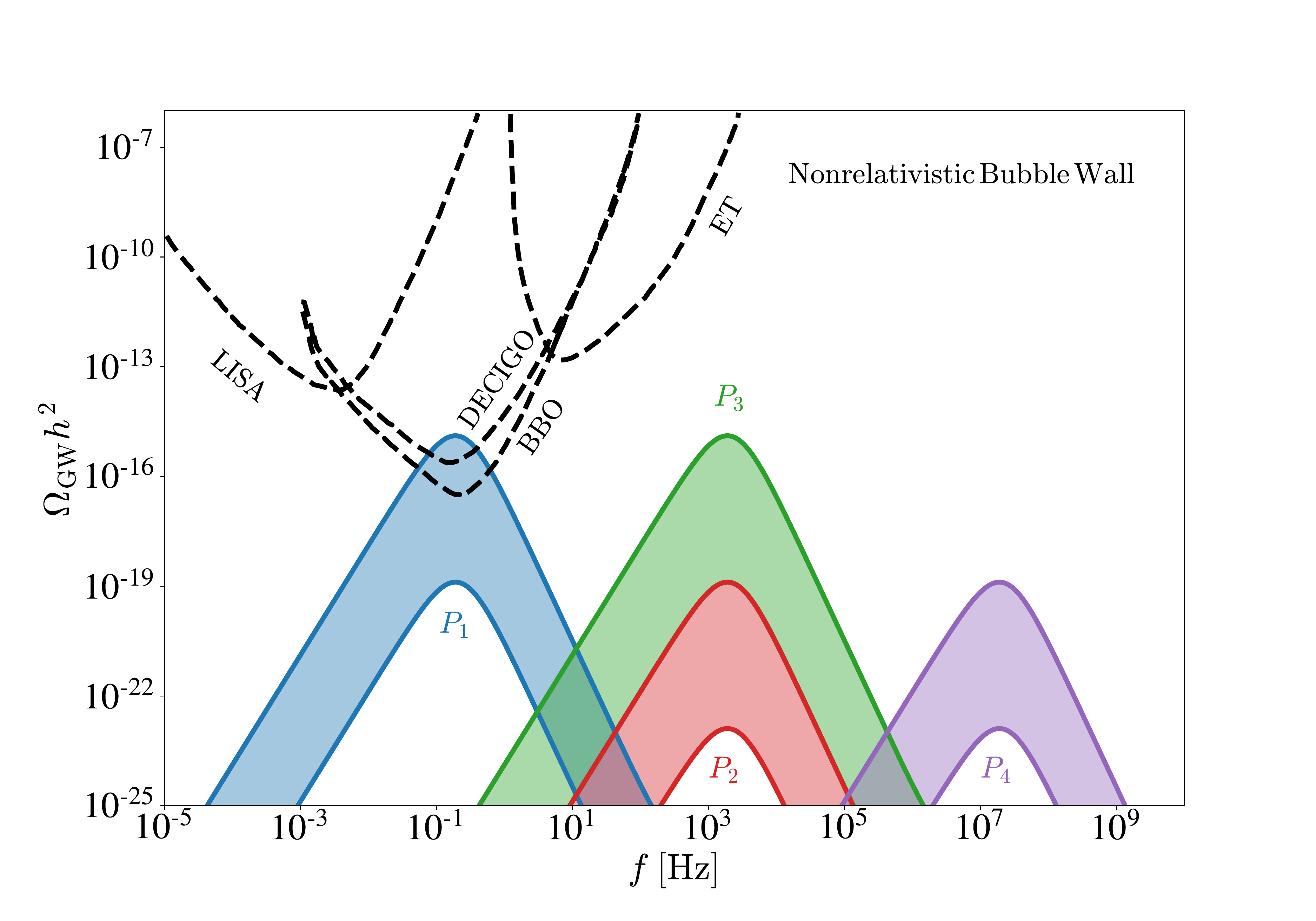}
    \includegraphics[width = .49\linewidth]{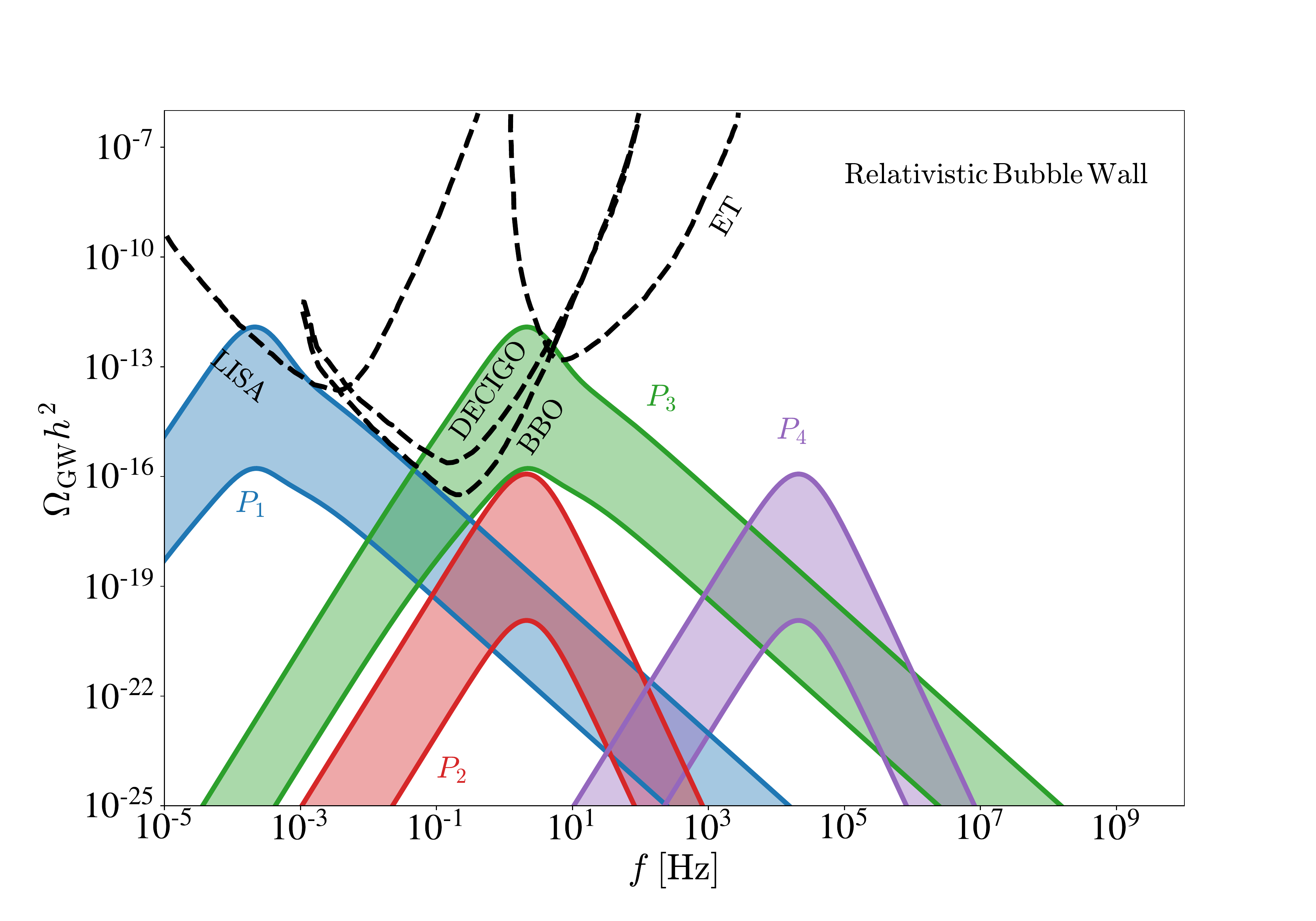}
    \caption{Spectrum of stochastic gravitational waves produced by a FOPT.
    The four bands represent phase transitions with: \( (T,\beta_H) = (10^3 \,\rm{GeV}, 1)\) (P1, blue), \((10^3 \,\rm{GeV}, 10^4)\) (P2, red),  
    \((10^7 \,\rm{GeV}, 1)\) (P3, green), and \((10^7 \,\rm{GeV}, 10^4)\) (P4, purple) with latent heat values \(\alpha \geq 0.01\). 
    The black dashed lines show projected experimental reaches from the future experiments LISA, DECIGO, BBO, and ET.}
    \label{fig:GW}
\end{figure*}

A first order phase transition produces a stochastic background of gravitational waves that could be observable.
The gravitational wave spectrum is thought to be composed of three primary components,
\begin{align}
    \Omega_{gw}h^2 = \Omega_{b}h^2 +\Omega_{s}h^2 +\Omega_{t}h^2 ,
\end{align}
where b, s, and t represent the bubble collision, sound wave, and turbulence contributions to the gravitational wave spectrum, and
whose contributions can be estimated in terms of quantities characterizing the properties of the transition itself,
 \(\alpha\), \(\beta_H \equiv \beta/H_*\), \(T_*\), \(v_w\), and \(\kappa_i\), where:
\begin{align}
    \beta_H = \left.\left(T \frac{\mathrm{d}}{\mathrm{d} T}\left(\frac{S_{3}(T)}{T}\right)\right)\right|_{T=T_{*}} .
\end{align}
These parameters can be estimated based on the properties of the phase transition, and mapped on to the
peak amplitudes, frequencies, and spectral shapes of the gravitational wave signal~\cite{Alanne:2019bsm},
with the spectral shape largely governed by \(v_w\), \(\beta_H\) and \(T\), whereas the amplitude is also sensitive 
to the latent heat, \(\alpha\) and efficiency factors, \(\kappa_i\). The amplitudes increase for large wall velocities making the relativistic bubble wall case more potentially
observable by future experiments.  

We sample all values of \(\alpha \geq 0.01\) since the amplitudes scale as \((\Omega h^2)^{\rm{peak}}\sim (\alpha/(1+\alpha))^n\) with \(n>0\),
and the amplitude asymptotically approaches its maximum as \(\alpha\gg 1\). 
We use typical values found in \cite{Alanne:2019bsm} for the efficiencies, \(\kappa_b = 10^{-8}\) and \(\kappa_v = 10^{-3}\). 
In Fig.~\ref{fig:GW} , we plot bands that represent the range of sampled \(\alpha\) values, 
and consider four benchmark parameter points for both relativistic and non-relativistic bubble walls
with \(\beta_H \sim 1, \,10^4\), \(v_w \sim 0.9,\, 10^{-3}\), and \(T = 10^3,\,10^7\) GeV.
Also shown are the projected sensitivities of the future GW experiments, LISA \cite{Auclair:2019wcv}, DECIGO \cite{Kudoh:2005as}, BBO \cite{Corbin:2005ny}, 
and ET \cite{Hild:2010id}. Gravitational wave signals generated in this scenario could be discovered for some of the parameter space considered, 
with phase transitions resulting in relativistic bubble expansion being more easily detectable. 

\section{Conclusions and Outlook}
\label{sec:conclusions}

We have explored a novel interplay between decay and annihilation that arises 
during a first-order phase transition
in  which the mass of the annihilating/decaying particle 
receives a large contribution from the phase transition.
We find that pocket collapse produces an enhancement of the particle density, and  opens up a large range of parameter space where annihilation can be important, 
which would typically otherwise be dominated by decay. We investigate baryogenesis during this type of scenario, where the decay and annihilation of \(\chi\) may both 
separately contribute to generating the observed baryon asymmetry.  While this is a novel application, the interplay between annihilation and decay during such a phase
transition is interesting in its own right, and may prove useful in other applications as well.

There are many interesting avenues for future exploration. For example, we approximate a constant temperature of the thermal bath throughout the phase transition,
but this need not be the case.  Indeed, the duration of the phase transition is longer than the Hubble scale for pockets whose sizes are initially \(\sim R_H\), 
such that the temperature of the universe may cool appreciably.  Other types of phase transition may themselves generate significant amounts of heating. 
We further assumed a constant bubble wall velocity, but depending on the heating during the phase transition, and the pressure exerted on the bubble wall by \(\chi\) 
could lead to a non-trivial wall velocity profile. Studying this more complicated evolution is left for future work. 

It would also be interesting to move beyond generic characterizations and see how these results could be applied to specific models of baryogenesis.
For example, \(\chi\) could be a right handed neutrino in a seesaw model of neutrino masses, whose large mass could be the result of the vacuum expectation
value of a field spontaneously breaking lepton number.  Typically in leptogenesis models, decays dominate and annihilation is negligible;
but for the an appropriate type of phase transition this expectation could be upset, leading to a different mapping between the phases of the
neutrino masses and Yukawa couplings and the resulting baryon asymmetry (from annihilating sterile neutrinos),
and violating the Davidson-Ibarra bound \cite{Davidson:2002qv}, or the need for tiny mass differences required by resonant leptogenesis  \cite{Pilaftsis:1997jf}. 

If the \(\chi\) is stable, it could play the role of dark matter, and it might be possible to generate the observed dark matter relic abundance and baryon asymmetry at the same time
\cite{Cui:2011qe,Cui:2011ab}.  Even without addressing baryogenesis, the enhancement of the annihilation could relax
the relationship between the annihilation cross section and the mass implied by freeze-out production of the dark matter, 
allowing small values of the cross section to generate the correct amount of dark matter.  We leave the investigation of these ideas for future work. 

\section*{Acknowledgements}

This work was supported in part by the NSF via grant number PHY-1915005.  

\bibliography{ref}

\end{document}